\documentclass[%
 reprint,
 amsmath,amssymb,
 aps,
]{revtex4-2}

\usepackage{graphicx}
\usepackage[usenames]{color}
\usepackage{dcolumn}
\usepackage{bm}

\begin{document}


\title{Scintillated microlensing: measuring cosmic distances with fast radio bursts}

\author{Anna Tsai}
\email{anna.tsai@mail.utoronto.ca}
\affiliation{Canadian Institute for Theoretical Astrophysics, University of Toronto, 60 St. George Street, Toronto, ON M5S 3H8, Canada}
\affiliation{Department of Physics, University of Toronto, 60 St. George Street, Toronto, ON M5S 1A7, Canada}
\author{Dylan L. Jow}%
 \email{djow@physics.utoronto.ca}
\affiliation{Canadian Institute for Theoretical Astrophysics, University of Toronto, 60 St. George Street, Toronto, ON M5S 3H8, Canada}
\affiliation{Department of Physics, University of Toronto, 60 St. George Street, Toronto, ON M5S 1A7, Canada}
\affiliation{Dunlap Institute for Astronomy \& Astrophysics, University of Toronto, AB 120-50 St. George Street, Toronto, ON M5S 3H4, Canada}

\author{Daniel Baker}
\affiliation{
Institute of Astronomy and Astrophysics, Academia Sinica, Astronomy-Mathematics Building, No. 1, Section 4,
Roosevelt Road, Taipei 10617, Taiwan
}%

\author{Ue-Li Pen}
\affiliation{
Institute of Astronomy and Astrophysics, Academia Sinica, Astronomy-Mathematics Building, No. 1, Section 4,
Roosevelt Road, Taipei 10617, Taiwan
}
\affiliation{Canadian Institute for Theoretical Astrophysics, University of Toronto, 60 St. George Street, Toronto, ON M5S 3H8, Canada}
\affiliation{Department of Physics, University of Toronto, 60 St. George Street, Toronto, ON M5S 1A7, Canada}
\affiliation{Dunlap Institute for Astronomy \& Astrophysics, University of Toronto, AB 120-50 St. George Street, Toronto, ON M5S 3H4, Canada}
\affiliation{Perimeter Institute for Theoretical Physics, 31 Caroline St. North, Waterloo, ON, Canada N2L 2Y5}
\affiliation{Canadian Institute for Advanced Research, CIFAR program in Gravitation and Cosmology}

\date{\today}

\begin{abstract}

We propose a novel means of directly measuring cosmological distances using scintillated microlensing of fast radio bursts (FRBs). In standard strong lensing measurements of cosmic expansion, the main source of systematic uncertainty lies in modeling the mass profile of galactic halos. Using extra-galactic stellar microlensing to measure the Hubble constant avoids this systematic uncertainty as the lens potential of microlenses depends only on a single parameter: the mass of the lens. FRBs, which may achieve nanosecond precision on lensing time delays, are well-suited to precision measurements of stellar microlensing, for which the time delays are on the order of milliseconds. However, typical angular separations between the microlensed images on the order of microarcseconds make the individual images impossible to spatially resolve with ground-based telescopes. We propose leveraging scintillation in the ISM to resolve the microlensed images, effectively turning the ISM into an astrophysical-scale interferometer. Using this technique, we estimate a 6\% uncertainty on $H_0$ from a single observed scintillated microlensing event, with a sub-percent uncertainty on $H_0$ achievable with only 30 such events. With an optical depth for stellar microlensing of $10^{-3}$, this may be achievable in the near future with upcoming FRB telescopes.

\end{abstract}


\maketitle


\section{Introduction}

The direct measurement of cosmological distances remains a fundamental challenge in cosmology. Strong lensing is one of the few methods capable of making direct measurements; however, traditional strong lensing techniques suffer from systematic uncertainties due to line-of-sight contamination and difficulty in modeling the lens mass profile. The appeal of strong lensing methods lies in their ability to provide $H_0$ measurements independent of Type Ia supernovae measurements and the cosmic microwave background (CMB): the main drivers of the so-called ``Hubble tension". Ref.~\cite{Wong_2019} uses galactic lensing of quasars to measure time delay distances, resulting in a Hubble constant of $73.3^{+1.7}_{-1.8}$ (km/s)/Mpc. This measurement is in agreement with local Type Ia supernova measurements, but in 3.1$\sigma$ tension with the $Planck$ CMB measurements ($H_0 = 68.20 \pm 0.63$ (km/s)/Mpc \cite{2020}). The distance measurements used to infer $H_0$ accumulate errors of between $2\%$ and $8\%$ from uncertainties in the lens-profile modelling and between $2.7\%$ and $6.4\%$ from line-of-sight effects \cite{Wong_2019}. We propose a new strong lensing method of measuring cosmological distance using stellar microlensed extra-galactic fast radio bursts (FRBs).

The term ``microlensing", here, refers to lensing by stellar-mass, compact objects whose lensing potential is unambiguously determined by the total mass, thereby avoiding the systematic uncertainties associated with modeling the complicated mass profiles of galactic halos. Fast radio bursts (FRBs) are extremely bright, short flashes of coherent radio emission at cosmological distances. Since FRBs are coherent sources of radiation, uncertainties in measurements of the lensing time-delays are limited by the wavelength of the light, potentially yielding nanosecond precision \cite{Wucknitz_2021}. This is a many-orders-of-magnitude improvement on the standard quasar time-delay precision of tens of days. In particular, this remarkable precision will make FRBs sensitive to stellar microlenses, which will have typical time delays of order milliseconds \cite{Jow_2020}. 

Time delays alone are not sufficient to derive a distance with which to measure cosmic expansion; a measurement of the angular separation of the microlensed images is also needed. However, typical angular separations for microlensed FRBs are on the order of a microarcsecond, well below the resolving power of even the largest baselines on earth. To resolve the angular separation between the microlensed images, we propose to use the scintillation of the lensed images due to multi-path propagation in the Milky Way interstellar medium (ISM) to effectively turn the ISM into an astrophysical-scale interferometer. A substantial fraction of FRBs are observed to scintillate \cite{Masui2015, Schoen2021}, and therefore, microlensed FRBs will also scintillate, making FRBs amenable to this technique. We estimate that a single scintillated microlensing event can achieve a $6\%$ uncertainty on $H_0$, and anticipate that enough observations of such lensing events are realistically obtainable in the next few years to achieve sub-percent uncertainty on $H_0$. A direct measurement of the Hubble constant to percent accuracy  would be competitive with other local probes of cosmic expansion and would inform debates on the Hubble tension \cite{Di_Valentino_2021}.


\section{\label{sec:level1}Microlensing}

\begin{figure}[b]
\includegraphics[scale = 0.31]{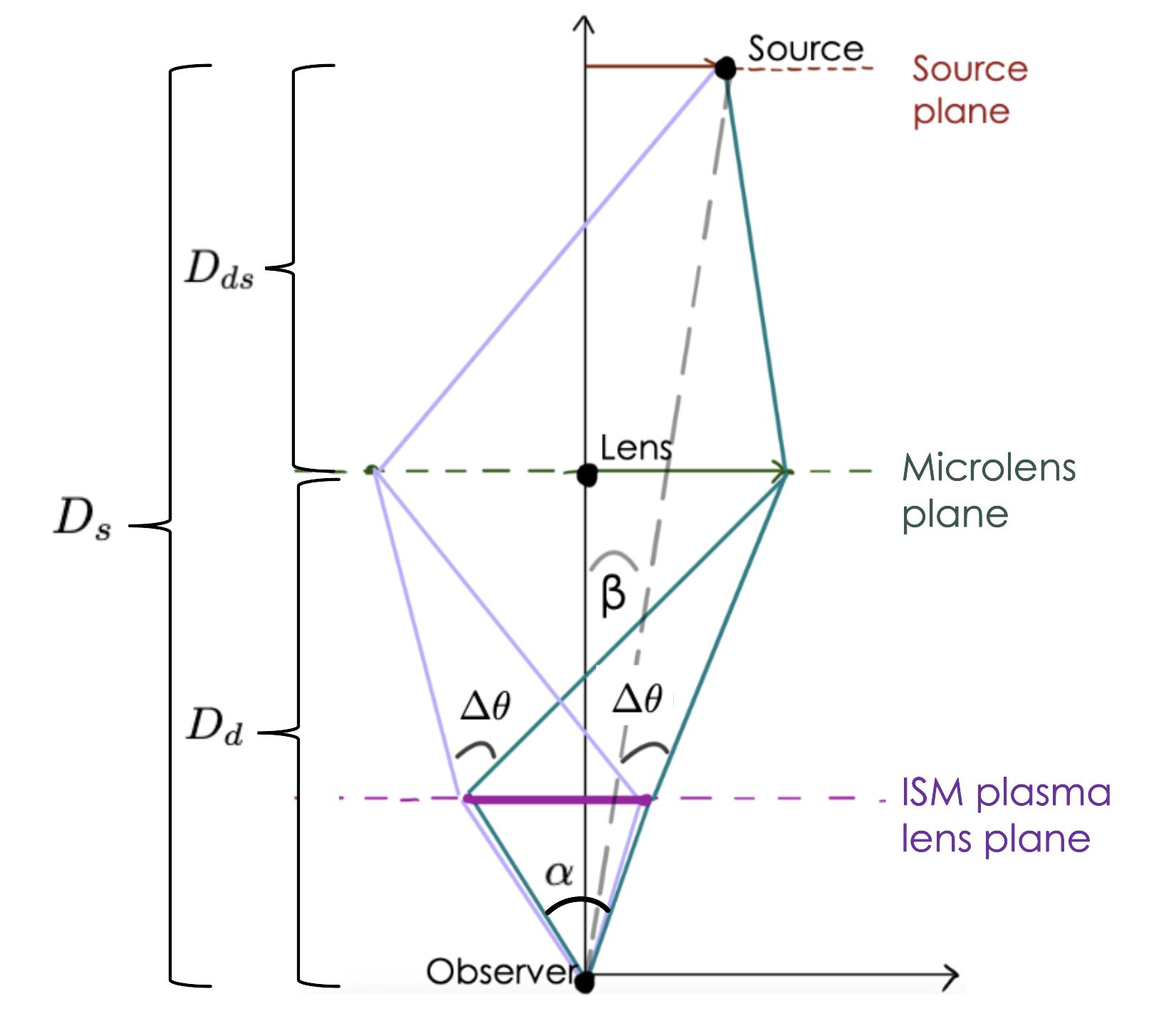}
\caption{\label{fig:lens} An FRB at a distance $D_s$ away from the observer, with an angular position given by the two-dimensional angle $\beta$ from the chosen optical axis, experiences microlensing from a star located at a distance $D_d$ from the observer. The microlensed images pass through a scintillating screen in the Milky Way ISM at a distance from the observer that is much smaller than the distances to the lens and source. The images formed by scattering in the ISM are separated by angle $\alpha$, whereas microlensed images are separated by angle $\Delta \theta$.}
\end{figure}

We consider a microlensed FRB that is also scintillated due to scattering in the ISM (Fig. \ref{fig:lens}). The Fermat potential for microlensing is given by \cite{Schneider_1992}:

\begin{equation} \label{eq:fermat}
    T(\boldsymbol{\theta}, \boldsymbol{\beta}) = \frac{D_d D_s}{2D_{ds}} |\boldsymbol{\theta - \beta}|^2 - 4GM \log|\boldsymbol{\theta}|,
\end{equation}
where $D_d$, $D_s$, and $D_{ds}$ denote distances shown in Fig.~\ref{fig:lens}, $\beta$ is the two dimensional angle between source and optical axis, $\theta$ is the two dimensional angle between optical axis and image, and $M$ is the microlens mass.

The angular position of the two microlensed images are given by points with stationary Fermat potential:
\begin{equation} \label{eq:stationary_pt}
    \boldsymbol{\theta_{\pm}} = \frac{\boldsymbol{\beta}}{2 \beta} \left(\beta \pm \sqrt{\frac{16 G M}{D} + \beta^2}\right),
\end{equation}
where $\beta = |\boldsymbol{\beta}|$ and $D \equiv \frac{D_d D_s}{D_{ds}}$, and the labels $\pm$ refer to the brighter and dimmer microlensed images, respectively.  For stellar microlenses, we are safely in the geometric optics regime \cite{Jow_2020}, and the effect of lensing is to magnify the two images by an amount
\begin{equation} \label{eq:mag}
    \mu_{\pm} = \frac{\beta + 8GM/D}{2\beta \sqrt{\beta^2 + 16GM/D}} \pm \frac{1}{2}.
\end{equation}

There are three observables that one can measure from a microlensing event: the relative time delay, angular separation, and magnification between the two images. We define these observables as follows:
\begin{equation} \label{eq:delta_T}
    \Delta T(D, \beta,M) \equiv T(\boldsymbol{\theta_+}) - T(\boldsymbol{\theta_-}), 
\end{equation}

\begin{equation} \label{eq:delta_theta}
    \Delta \theta(D, \beta, M) \equiv |\boldsymbol{\theta_+} - \boldsymbol{\theta_-}|,
\end{equation}

\begin{equation} \label{eq:rho}
    \rho(D, \beta, M) \equiv \frac{\mu_+}{\mu_-}.
\end{equation}
The effective distance, $D$, is uniquely determined by these three observables, and is given by:
\begin{equation}
    D = \frac{4G}{\Delta \theta ^2}  \left(\frac{(4+y^2) \Delta T}{8G \log(y - \sqrt{4+y^2}/2)-2G(y \sqrt{4+y^2})}\right),
\end{equation}
where $y^2 = \frac{\rho + 1}{\sqrt{\rho}} - 2$. Thus, measuring the three observables allows one to infer the effective distance, $D$. For stellar mass microlenses, the time delays are of order $\Delta T \sim 1\,{\rm ms}$, independent of distance, well within the prescision achievable with FRBs. However, for stellar masses and cosmological distances ($D \sim 1\,$Gpc), the angular separation is on the order of $\Delta \theta \sim 1\,\mu$as, which is far smaller than typical angular resolutions achievable by ground-based radio telescopes such as the Canadian Hydrogen Intensity Mapping Experiment (CHIME). Even with the upcoming addition of very-long baseline interferometry (VLBI) outriggers, the CHIME-TONE array will have a maximum baseline of $\sim 3300$ km\,\cite{CHIMETONE2023} which, at $1$ GHz, can only resolve $30$\,mas. In the following section, we propose leveraging scintillation of microlensed FRBs in the ISM as a way to resolve the angular separation of the microlensed images. 

\section{Scintillating Screens in the ISM as VLBI}

Bright sources of coherent radio emission are observed to scintillate due to multi-path propagation in the ISM. Recent VLBI imaging of scintillating pulsars reveal that the scattered images are co-linear on the sky, with an aspect ratio of within one percent \cite{Brisken_2009} \cite{baker2022high}. To achieve co-linearity in the images to this degree, the scattering structures in the ISM must be highly anisotropic. In particular, a single, thin screen of effectively one-dimensional inhomogeneities must dominate contributions to the total bending angle. Such screens have been found to accommodate a wide range of pulsar scintillation observations \cite{Brisken_2009, PenLevin2014, baker2022high, Jow_2023}.

Just as ground-based VLBI can be used to resolve the angular separation of scattered images of scintillating pulsars, we propose to use the scattered images of scintillating FRBs as an effective astrophysical-scale interferometer to resolve the much smaller angular separations due to microlensing.
Scintillation in the ISM is observed to produce hundreds of images on the sky, with $b \sim 10\,{\rm AU}$ separations in the plane of the ISM scattering screen. These enormous baselines can achieve angular resolutions of $\lambda / b \sim 10^{-2}\,\mu{\rm as}$. By using ground-based VLBI to resolve the ISM-scattered images of scintillating FRBs, we can leverage the effective baselines provided by scattering in the ISM to achieve resolutions many orders-of-magnitude finer than what can be achieved with ground-based VLBI. While this technique almost seems too good to be true, it has been successfully implemented to resolve micro- and pico-arcsecond features in pulsar emission regions \cite{Pen_2014, Lin_2023}.  However, in contrast to pulsar scintillation techniques, FRB measurements cannot relay on time domain information. Whereas pulsars continuously emit bursts, allowing measurements of time dependent modulation as the earth sweeps through the induced scintillation pattern, non-repeating FRBs produce single bursts of millisecond duration. In principle, the same information can still be obtained from an instantaneous FRB signal if the signal is observed simultaneously at multiple stations around the earth. It is unknown the necessary density and number of arrays across the earth that are needed to recover the timing information necessary to apply pulsar scintillation arch techniques. A systematic investigation of this issue, although necessary, is beyond the scope of the present letter and will be the focus of future work. Here we will attempt to demonstrate the principle of proposed measurement for FRB microlensing.

\begin{figure}[b]
\includegraphics[scale = 0.45]{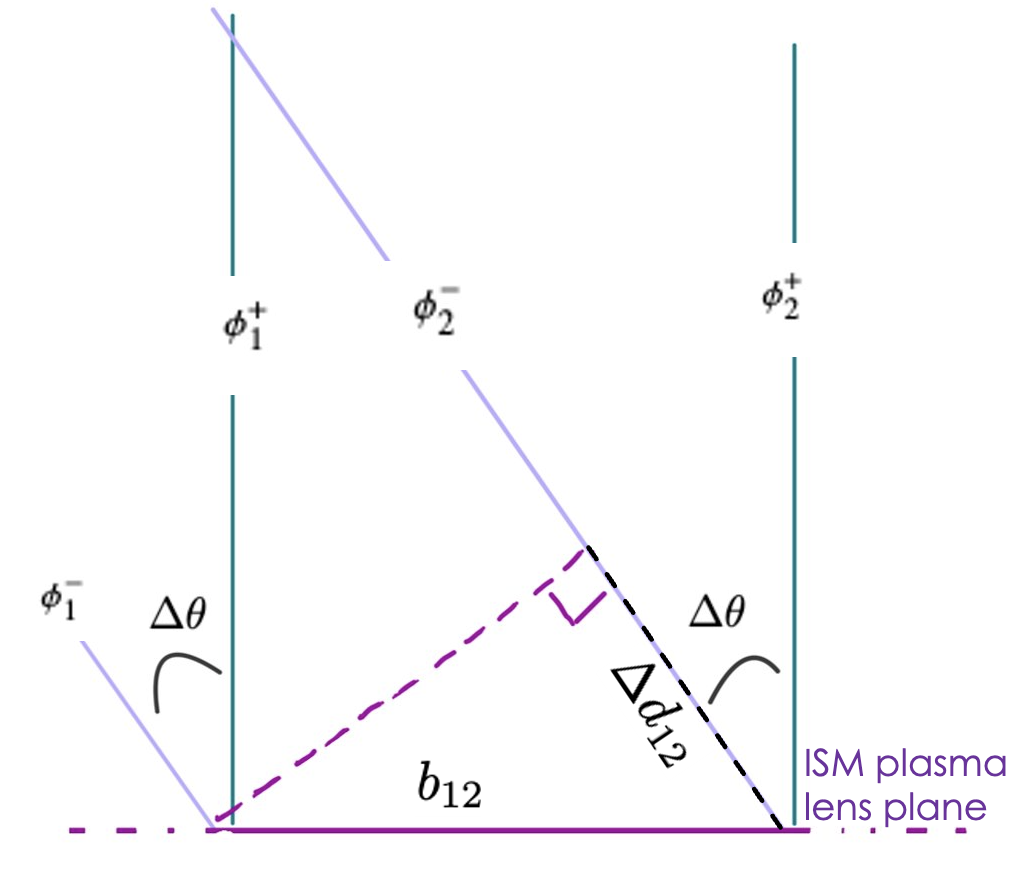}
\caption{\label{fig:ISM} Four rays pass through a single ISM plasma screen with phases labeled by $\phi$.  Superscripts (+/-) on the phase correspond different microlensed images.  Numeric subscripts denote scattered images.  Rays marked by the same superscript are assumed to be parallel.}
\end{figure}

Consider a microlensed FRB whose rays passing through a single scattering screen in the ISM, using the simple case of only two scattered images. In reality, the ISM screen will result in hundreds of scattered images, but, for the sake of clarity, we will use this simple picture to illustrate the technique and note that a more complete analysis can be carried out in practice. For two scattered images, there is a single angular separation, $\alpha$ (shown in Fig.~\ref{fig:lens}), which we assume can be measured with ground-based VLBI. Moreover, using standard techniques from pulsar scintillation, we assume that the distance to the ISM screen can be inferred from ground-based VLBI.  Using estimates from \cite{baker2022high}, typical values of $\alpha$ will be on the order of $\sim 10$mas and the distance to the scintillating screen will be $\sim\,{\rm kpc}$, yielding a baseline on the ISM plane of $b_{12}\sim 10\,$AU. Using the ISM screen as an effective interferometer, one is sensitive to the combined phase offset, $(\phi^-_2 -\phi^+_2) - (\phi^-_1 -\phi^+_1)$, where $\phi^{\pm}_j$ is the phase of the ray from the microlensed image, $\theta_\pm$, incident on the ISM at the location of the $j^{th}$ scattered image (see Fig.~\ref{fig:ISM}). The technical details of how one can infer this combined phase from scintillation are described in detail elsewhere \cite{Pen_2014,Lin_2023}. We would like to relate the combined phase offset to the angular separation between the two microlensed images so that the latter can be inferred from the former.

To proceed, we will make a few simplifying assumptions.  Firstly, the distance to the scintillating screen, $\sim\,$kpc, is negligible in comparison to $D_s$ and $D$ (both $\sim\,$Gpc). Secondly, the rays incident on the ISM screen from the same microlensed image (i.e. the rays with the same numeric subscript in Fig.~\ref{fig:ISM}) are effectively parallel when they hit the ISM screen.  This assumption is true in the limit as the microlenisng plane becomes infinitely far away, and, thus, follows from the first assumption. Lastly, we assume that the phase offset between the rays is purely geometric. That is, we neglect the contribution to the phase from the refractive index of the ISM and the gravitational potential of the microlens. Indeed, given the vast distances traversed by the rays, the geometric terms in the time delay typically dominate in both microlensing and scintillation.

With these assumptions in hand, we can now relate a combination of phase differences to the angular separation between microlensed images ($\Delta \theta$):

\begin{equation} \label{eq:phase_theta}
    (\phi^-_2 - \phi^+_2) - (\phi^-_1 - \phi^+_1) = \frac{\Delta d_{12}}{\lambda} = \frac{\Delta \theta b_{12} 2 \pi}{\lambda}.
\end{equation}
The effective baseline, $b_{12}$, can be measured using ground-based VLBI, thus allowing Eq.~\ref{eq:phase_theta} to be inverted to find the angular separation of the microlensed images.


\section{Error on $H_0$}

The main source of statistical error in a measurement of $H_0$ will come from the uncertainty in the measurement of the relative phase offsets. For bursts with high signal-to-noise, the uncertainty in measurements of the magnification ratio, $\rho$, will be small in comparison. Of the three microlensing observables, the angular separation and $\Delta \theta$ depend on phase measurements by Eq \ref{eq:phase_theta} and $\Delta T$ depends on the phase offset by $(\phi^- - \phi^+) = 2 \pi \nu \Delta T$.  Using a conservative estimate of instrumental error, one may realistic attain an uncertainty in the phase, $\phi^- - \phi^+$, of $3\%$ \cite{baker2022high}, which yields a $6\%$ uncertainty on the inferred distance, $D$. 

The effective distance, $D = D_d D_s / D_{ds}$, is a combination of distances. In order to measure $H_0$, we need to be able to measure the lens distance, $D_d$, as a function of redshift. We will assume that the FRB is well-localized so that a host and lens galaxy, and thereby, a host and lens redshift, can be inferred. Since we already require that the lensing event in question be observed using ground-based VLBI, any such event will automatically be well-localized. With the source and lens redshifts, one can infer $H_0$ from the effective distance \cite{Suyu_2017}. To obtain a rough estimate on the achievable uncertainty in $H_0$, we will consider the low-redshift regime where $z_d = D_d H_0 /c$, independent of other cosmological parameters. In this regime, a single scintillated microlensing event will yield a measurement of $H_0$ with a $6\%$ uncertainty. The uncertainty scales with the number of observed lensing events like $\sigma_{H0} \propto \frac{1}{\sqrt{N}}$. Thus, with approximately 30 observed events, we can measure $H_0$ to within a $1\%$ uncertainty.

The optical depth for microlensing by extra-galactic stars is roughly $\tau \approx 10^{-3}$ \citep{Connor2023}.  This yields an event rate for CHIME (with a rate of $\sim$10 FRB detections per day) of approximately 4 stellar microlensed FRBs per year. This event rate will likely increase, potentially by up to two orders of magnitude, as radio telescopes geared towards high event-rates are built. For instance, the proposed Packed Utlra-wideband Mapping Array (PUMA) anticipates an FRB detection rate of more than one-thousand per day \cite{PUMA2019}. Moreover, higher sensitivity telescopes will extend sensitivity to lower-magnification microlensing events, thereby simultaneously increasing the optical depth of microlensing. With many next-generation FRB telescopes set to come online in the coming years \cite{CHORD2019, BURSTT2022}, there will be an abundance of FRB data with which to search for scintillated microlensing events. We know from pulsar scintillation that practically every sight-line through the ISM results in scintillation; however, not every sight-line will be appropriate for our proposed method, as we require the scintillation to be dominated by a single anisotropic screen. Extrapolating from a recent pulsar scintillation survey \cite{Stinebring_2022}, we expect that a substantial percent of sight-lines scintillate in the required way, and make a very rough estimate of $50\%$. This estimate is also the subject of considerable uncertainty in our analysis, as the survey in Ref.~\cite{Stinebring_2022} is by no means an unbiased sample of pulsars. However, it is clear from this and other detections \cite{Masui2015, Hessels, Marcote, Wu, Sammons} of FRB scintillation that anisotropic scintillation with the necessary properties are not extremely rare occurences for radio sources.
Even with $\sim$4 lensing events a year (using the current CHIME detection rate), observing a total of thirty scintillated microlensing events in the coming years is not unrealistic.

\section{Systematic Uncertainties}

Although microlenses are potentially much cleaner systems than the complicated halo profiles of standard strong lensing measurements, they are not totally without systematic effects that may bias parameter estimation. Here we discuss potential sources of systematic uncertainty. 

\subsection{Angular broadening}
Angular broadening of FRB emission due to scattering by density irregularities in both the FRB host galaxy and in the region of the stellar microlens is potentially a major source of systematic uncertainty.  Scintillation depends on the point-like nature of FRB emission, and if angular broadening is significant, the FRB signal will no longer scintillate.  With regard to angular broadening by scattering in the FRB host galaxy, it is difficult to make a precise estimate on the significance of this effect in the absence of an unbiased survey of FRB scintillation.  However, it is clear from several studies \cite[see e.g.][]{Masui2015, Hessels, Marcote, Wu, Sammons} that many FRBs do in fact strongly scintillate in the ISM.  Thus, there is at least a population of FRBs for which angular broadening by the FRB host galaxy is a small enough effect that the FRB signal still scintillates in the ISM. Angular broadening due to scattering near the stellar microlens, however, remains a genuine issue for our proposed measurement. While it is at least in principle be possible to know when a stellar microlensed FRB signal is also contaminated by scattering in the lens' hos galaxy (since plasma scattering is a chromatic effect, whereas gravitational lensing is achromatic), the fraction of events that will be so contaminated is unknown both empirically and theoretically, as the plasma structures on the relevant scales are highly unconstrained \cite{VP2019,JowWu2023}. 

\subsection{Binary contamination}
\label{sec:binary}

Some fraction of stellar systems are binaries. While the exact fraction will depend on the lens' host galaxy, the fraction is typically larger than ten-percent and potentially as large as one half. Falsely identifying binary lenses as point-like microlenses may introduce bias in the parameter estimation. Binary lenses differ from point lenses primarily by the presence of extended caustics. The caustic structure of a binary lens is governed only by the ratio of the two masses, $q = M_1 / M_2$, and the angular separation between the masses, $s = \Delta \theta / \theta_E$, normalized by the Einstein angle for the total mass, $\theta_E = \sqrt{4 G (M_1 + M_2) / D}$. For typical stellar binaries, the mass ratio is of order one. At cosmological distances, a separation of $\sim 1\,{\rm AU}$ between the masses yields an angular separation of $\Delta \theta \sim 0.001\,\mu{\rm as}$. A typical Einstein radius for a solar mass system at cosmological distance is $\theta_E \sim 1\,\mu{\rm as}$, so that $s \sim 10^{-3}$. Note that the separation parameter scales inversely with the distance to the lens, $s \sim D^{-1/2}_d$, so that $s$ will always be much less than unity for cosmological stellar lensing. For such small values of $s$, the binary lens will have the ``close" caustic topology (see Ref.~\cite{Gaudi2012} for a classification of binary-lens caustics). The primary caustic is a diamond shape, centred around the optical axis, contained within the Einstein ring. For a binary lens with $q = 1$ and $s = 10^{-3}$, the angular area of the caustic is $\sim 10^{-12} \theta_E^2$. In other words, the optical depth of the binary caustics at cosmological distances is negligible. The observables for the binary lens only differ substantially from an equivalent-mass microlens within this caustic region. Indeed, at an impact parameter of $10^{-3} \theta_E$ from the optical axis, the magnification ratio between the brightest images differs from the microlensing predictions by less than one percent, and the relative time delay differs by less than $10^{-8}\%$. While, in principle, for the binary lens there is always an additional third image that does not exist for the microlens, outside of the central caustic this image is typically below detection thresholds. Moreover, when more than two lensed images are detected, such events can be excluded as non-microlensing events. 

\subsection{Shear}
\label{sec:shear}

Stellar microlenses microlensing events will be affected by the shear from the lens' host galaxy halo. In principle, if the FRB can be localized, and the galaxy containing the microlens is known, then the shear can be estimated, and its effect fully accounted for via the Chang-Refsdal lens model \citep{ChangRefsdal1979, 2006MNRAS.369..317A}. In general, however, the external shear of a galaxy halo at cosmological distances is small. For a source at redshift $z_s = 1$, and an NFW lens halfway between the source and observer with concentration parameter $c_{200} = 10$ and virial radius $r_{200} = 100\,{\rm kpc}$, the shear within the characteristic radius is $\gamma \sim 0.01$, assuming standard cosmological parameters \citep{Wright2000}. For shears of this magnitude, the effect on the bending angle, and hence the magnifications, will be small. The main effect will be on the inferred time delays. The maximum additional time delay due to the shear occurs when the microlensed images are oriented along the direction of the shear, and is on the order of $\tau_{\gamma} \sim \frac{\gamma}{2} \frac{D}{2c} \theta^2_E \sim \gamma \Delta T$. Thus, the effect of shear on the observed time delay will typically be below the percent level. 

\subsection{Line of sight effects}
\label{sec:los}

Mass inhomogeneities along the line of sight contribute up to $6.4\%$ uncertainty on each lensing observation in previous strong lensing studies \cite{Wong_2019}. However, for stellar microlensing, line-of-sight effects are mitigated by the small angular separation of the lens images. In other words, both images take effectively the same path from source to observer, and are affected by matter along the line of sight in the same way. Since we are only sensitive to differences in arrival times between the two images, stellar microlenisng is largely insensitive to line-of-sight inhomogeneities. Compared to standard macrolensing measurements, for which arcsecond angular separations lead to differences in path length of order $\sim\,10\,{\rm kpc}$, path length differences for the two microlensed images will be $\sim100\,$AU.

\section{conclusion}

We propose a new strong lensing method for measuring the Hubble constant which employs stellar microlensed FRBs that are also scintillated. Scintillation allows for much finer angular resolution than would be possible with ground-based telescopes. The primary advantage of using microlenses, over the galactic halo ``macrolenses" of standard strong lensing measurements, is that the lens potential of a microlens depends only a single parameter (the mass), and therefore does not have any systematic uncertainty related to the mass profile modelling. Using a conservative estimate for instrumental uncertainty in phase measurements, the proposed microlensing method can yield a measurement of $H_0$ to within $1\%$ uncertainty with approximately 30 observed events.

\nocite{*}

\bibliography{apssamp}

\end{document}